\documentclass[aps,floatfix,prb,twocolumn,a4paper,10pt,showpacs,citeautoscript,superscriptaddress,preprintnumbers]{revtex4-1}
\usepackage[utf8]{inputenc}
\usepackage[OT1]{fontenc}
\usepackage{amsmath}
\usepackage{amsfonts}
\usepackage{amssymb}
\usepackage{graphicx}
\usepackage{epstopdf}
\usepackage{epsfig}
\usepackage{subfigure}
\usepackage{stmaryrd}
\usepackage{ulem}
\usepackage{hyperref}

\begin{document}
\title{Thomas precession, persistent spin currents and quantum forces}
\author{S.Ouvry}
\affiliation{LPTMS, CNRS UMR 8626, Facult\'e des Sciences d'Orsay, Orsay 91405 France}
\author{L.Pastur}
\author{A.Yanovsky}
\affiliation{B. Verkin ILTPE,  NAS of Ukraine, 47 Lenin Ave, Kharkov 61103 Ukraine}

\begin{abstract}
We consider T-invariant spin currents induced by  spin-orbit interactions which originate from the confined motion  of  spin carriers in nanostructures. The resulting  Thomas  spin precession is a fundamental and purely kinematic relativistic effect occurring when the acceleration of carriers is not parallel to their velocity. In the case, where the carriers (e.g. electrons) have magnetic moment the forces due to the electric field of the spin current can, in certain conditions, exceed the van der Waals-Casimir forces by several orders of magnitude.
We also discuss a possible experimental set-up tailored to use these forces for checking the  existence of  a nonzero anomalous
magnetic moment of the photon.
\end{abstract}


\maketitle

\section{Introduction}
The studies of the electronic spin degrees of freedom,  spintronics, is an active branch of
solid state physics. In particular, spintronics of nanostructures, or nanospinstronics,
has developed quite rapidly in recent years \cite{Fert,Seneor,Kontos},  motivated  by a number of basic questions on the nature of nanophenomena as well as by
its potential impact on information technology. The  underlying physical mechanism
is the spin-orbit interaction of conducting electrons, which couples their spin degree of freedom to their orbital dynamics.

One distinguishes the extrinsic phenomena, the result of  spin dependent scatterings on (external) impurities, from the intrinsic phenomena, arising because of a certain spin-orbit coupled band structure. There are two basic dynamical mechanisms underlying intrinsic phenomena in semiconductors:
(i) the Rashba\cite{Rashba} coupling due to the combined effects of atomic spin-orbit coupling and structured inversion asymmetry;
(ii) the Dresselhaus\cite{Dresselhaus} coupling due to the bulk crystal inversion asymmetry.

The intrinsic spin-orbit interactions are essential for many potential applications like spin polarized currents without magnetism\cite{Awschalom}, spin field-effect transistor\cite{DataDas}, topological insulator states\cite{Kane1,Kane2}, etc.
One of the important spin-orbit effects in nanostructures is the persistent spin currents arising from the above dynamical sources of asymmetry, see e.g. \cite{Sun,Maiti}.

In this article we consider yet another mechanism for spin-orbit coupling whose origin is the Thomas spin precession \cite{ThomasNature1926} --\cite{Te-Bo:80}, a fundamental relativistic effect. It occurs when the particles acceleration is not parallel to their velocity, i.e. for any motion of  relativistic particles with curving or winding trajectories. The  precession has a purely kinematical origin since it  results solely from the confined motion of the particles in a sufficiently small volume.  It is not necessary related to  impurity scatterings or other   extrinsic\cite{RashbaExtrSpinHall,DasSarmaExtrSpinHall,SternExtrSpinHall} spin-orbit mechanisms. Likewise, the Thomas precession can occur even for ideal metallic nanoparticles   without Ras\-h\-ba or Dres\-sel\-ha\-us couplings.  As we will show below, the Thomas precession will  induce persistent spin currents whose dependence on the spin degree of freedoms and on the quantum spectrum is different from those resulting from the Ra\-sh\-ba-Dres\-sel\-ha\-us couplings.

The inverse asymmetry (the broken $P$-invariance) is the cause of spin persistent currents due to  spin-orbital coupling. However, time reversal invariance ($T$-invariance) is not broken in this case. This is in complete agreement with the symmetry of   pure spin (or other rotation) equilibrium currents, which are obviously not $P$-invariant but are $T$-invariant. Indeed, any geometric constraint breaks $P$-invariance.

A confining strip as an example of geometry in which there exist the winding persistent spin currents due to the Thomas precession has been already considered in Ref.\cite{Ouvry1} in the case of classical Brownian motion.

In this paper we discuss a quantum phenomenon in nanostructures, whose origin is also the Thomas precession. In this case the $P$-invariance with respect to winding and spin is also absent because of the confined motion of carriers.

\section{Thomas precession in nanostructures}
We consider an ideal and neutral metallic wire as an example of Thomas precession due to  the confinement of particles. Clearly if the wire were straight, the  motion of  free electrons along the wire  would have no relation to  precession and  spin-orbit interactions. If, however, the wire is curved into a closed loop, then the  motion of electrons along the loop will inevitably produce the Thomas precession of their spins.
Note that there is no electric field in this case and  the wire is neutral and equipotential.

Let us estimate the order of magnitude of spin-orbit effects resulting from  Thomas precession in a loop of nanosize assuming for simplicity that the loop is a ring. In the leading relativistic approximation the spin-orbit energy from Thomas precession is\cite{ThomasNature1926,jackson}
\begin{equation}
E_{s-o} \approx \frac{\vec{s}\cdot(\vec{v} \times \vec{\dot{v}})}{2 c^2} = \frac{v^2}{2 c^2} \vec{s}\cdot \vec{\dot{\theta}} \ .
\label{eq:e-so-thomas}
\end{equation}
Here the over-dot denotes  the time derivative, $\times$  the vector product, $c$ the speed of light, $\vec{s}$  the particle spin in angular momentum units, $\vec{v}$  its velocity and $\vec{\theta}$  its winding angle defined as
$$
\vec{\theta} = \int_0^t \frac{\vec{v} \times \vec{\dot{v}}}{v^2} dt \ .
$$
It is the vector sum of all the
windings of the particle trajectory (see \cite{Ouvry1}). The acceleration of the winding modes in a ring at constant angular velocity is simply $\dot{v} = v^2/r$ where $r$ is the radius of the ring.  Comparing   (\ref{eq:e-so-thomas}) with the commonly used estimate $ e \vec{s}\cdot(\vec{v} \times \vec{\mathcal{E}}_{eff})/{2 m c^2}$ of the spin-orbit energy  in an electric field $\vec{\mathcal{E}}_{eff}$,  where $e$ and $m$ are respectively the charge and the mass of the particle\cite{Bj-Dr:64}, we conclude that the Thomas precession of a free electron inside a  metallic nanoring
corresponds to a velocity dependent effective electric field
\begin{equation}\label{electric}
\mathcal{E}_{eff} \sim \frac{m v^2}{ e r} \ .
\end{equation}
Typically for electrons in metals the Fermi energy $\varepsilon_F$ is of the order of several electronvolts (e.g. $\varepsilon_F \sim 6$ eV for Au), hence the Fermi velocity is $v_F \sim  10^6$ m/s. Thus, in the case of nanoring, where $r \sim 10^{-9} \div 10^{-8}$~m and the particle velocity is of the order of the Fermi velocity $v\sim v_F$, the effective electric field in (\ref{electric}) can be quite large, i.e.,  $\mathcal{E}_{eff} \sim 10^{10}$ V/m. We conclude that the  spin-orbit effects can be well pronounced in nanorings. Note again that the electric field is just an "effective" field arising from the trajectory windings and there is no gradient of an electrical potential along a metallic nanoring.

Let us now show that the spin-orbit coupling (\ref{eq:e-so-thomas}) due to Thomas precession leads to persistent spin currents.
Note that we are using here a 1d ring and non interacting fermions  to obtain simple estimates, keeping in mind that we are interested in 2d or 3d nanostructures. 

The effective classical Lagrangian of free electrons which takes into account the Thomas precession is \cite{Frenkel,Te-Bo:80}
\begin{equation}\label{rel}
L = -m c^2 \sqrt{1 - \frac{v^2}{c^2}} - \vec{s} \cdot \left(\frac{1}{\sqrt{1 - \frac{v^2}{c^2}}} - 1 \right) \frac{\vec{v} \times \dot{\vec{v}}}{v^2}
\end{equation}
Note that the low velocity expansion of the second term in the r.h.s of (\ref{rel}) rightly reproduces (\ref{eq:e-so-thomas}).
In the  simple ring geometry we have in general
\begin{subequations}
\begin{align}
\vec{v} &= r \dot{\theta} \vec{e}_\theta
\\
\dot{\vec{v}} &= -r^{-1} v^2\vec{e}_{r} + r \ddot{\theta} \vec{e}_\theta \ ,
\label{eq:dot-dot-theta}
\end{align}
\end{subequations}
where $\theta$ is the polar angle, $\vec{e}_{\theta}$ and $\vec{e}_{r}$  are the polar unit vectors in the plane of the ring. It follows that for the ring geometry
\begin{equation} \frac{\vec{v} \times \vec{\dot{v}}}{v^2}=\dot{\theta}\vec{e}_z
\end{equation}
where $\vec{e}_z$ is the unit vector of the $z$-axis perpendicular to the ring, so that the winding angle and the polar angle are equal, thus denoted by the same symbol $\theta$). We  obtain  from (\ref{rel})
\begin{equation}\label{eq:no2dots}
L = -m c^2 \sqrt{1 - \frac{r^2 \dot{\theta}^2}{c^2}} - \left( \frac{1}{\sqrt{1 - \frac{r^2 \dot{\theta}^2}{c^2}}} - 1 \right) s_z \dot{\theta}
\end{equation}
where $ s_z$ is the spin projection on $\vec{e}_z$.
Note that the double derivative $\ddot{\theta}$ in (\ref{eq:dot-dot-theta})   does not contribute to the r.h.s. of (\ref{eq:no2dots}). Nevertheless the quantization of  (\ref{eq:no2dots}) is  quite non-trivial.  Fortunately, in the case of non interacting fermions, we can neglect states above the Fermi energy assuming  for simplicity zero temperature, i.e. $r\dot{\theta} = v \leq v_F << c$. Thus, expanding the r.h.s. of (\ref{eq:no2dots}) in the small parameter $v_F/c \sim 10^{-2}$ and dropping the constant energy $-mc^2$, we obtain
\begin{equation}
L \approx \frac{m r^2 \dot{\theta}^2}{2} - s_z \frac{r^2\dot{\theta}^3}{2 c^2},
\quad r \dot{\theta} \ll c.
\label{eq:class-exp}
\end{equation}

The angular momentum $p_\theta$ plays for (\ref{eq:no2dots}) the role of a generalized momentum
\begin{equation}
p_\theta = \frac{\partial L}{\partial \dot{\theta}} \approx m r^2 \dot{\theta} - \frac{3}{2} s_z  \frac{r^2 \dot{\theta}^2}{c^2}.
\end{equation}
Taking into account that $p_\theta \ll mrc \ll m^2 r^2 c^2 / \hbar$ (the second inequality means that the quantum uncertainty of the velocity  $\hbar/ m r$ has to be much smaller than $c$), i.e.,  $\dot{\theta} \ll c/r \ll m c^2 / \hbar$,
we can express  $\dot{\theta}$ as
\begin{equation}
\dot{\theta} \approx \frac{p_\theta}{m r^2} + \frac{3}{2} \frac{s_z}{m^3 r^4 c^2} p_\theta^2
\label{eq:dot-theta}
\end{equation}
Thus, using  (\ref{eq:class-exp}) and (\ref{eq:dot-theta}), the Hamiltonian corresponding to (\ref{eq:no2dots}) is
\begin{equation}
H = p_\theta \dot{\theta} - L \approx \frac{p_\theta^2}{2 m r^2} + \frac{1}{2}\frac{s_z p_\theta^3}{m^3 r^4 c^2}
\label{eq:hamiltonian-def}
\end{equation}
Now  the standard quantization procedure
\begin{equation}
p_\theta \rightarrow \hat{p}_\theta = - i \hbar \frac{\partial}{\partial \theta} \ , \ \ s_z \rightarrow \hat{s}_z = \frac{\hbar}{2} \hat{\sigma}_z.
\label{eq:p-operator}
\end{equation}
 yields the quantum Hamiltonian
\begin{equation}
\hat{H} = - \frac{\hbar^2}{2 m r^2} \frac{\partial^2}{\partial \theta^2} + i \frac{1}{4}\frac{\hbar^4}{m^3 r^4 c^2} \hat{\sigma}_z \frac{\partial^3}{\partial \theta^3} \ .
\label{eq:H-partial}
\end{equation}
Its spectrum has  doubly degenerated energy levels
\begin{align}\label{eq:spectrum}
\epsilon_{n\sigma_z} = \frac{\hbar^2 n^2}{2 m r^2} + \frac{\hbar^4 \sigma_z n^3}{4 m^3 r^4 c^2},
\\ \sigma_z = \pm1, \ n=0, \pm 1, \pm 2, ..., \notag
\end{align}
since $\epsilon_{n,\sigma_z}=\epsilon_{-n,-\sigma_z}$, and two-component states
\begin{equation}
\Psi_{n,\sigma_z} =\frac{1}{4\pi}e^{in\theta}\begin{pmatrix}1+\sigma_z\\1-\sigma_z\end{pmatrix}.
\label{eq:H-states}
\end{equation}
Eqs.~(\ref{eq:spectrum})-(\ref{eq:H-states}) can be used only for the  winding numbers $n$ satisfying  (cf. (\ref{eq:class-exp}) and (\ref{eq:dot-theta}))
\begin{equation}
|n| < N \ll \frac{m c r}{\hbar},
\label{eq:validity-spect}
\end{equation}
where
\begin{equation}
N \approx \frac{r \sqrt{2 m \varepsilon_F} }{\hbar}
\label{eq:fermilevel}
\end{equation}
is a natural cutoff fixed by the Fermi energy, i.e.  $\epsilon_{N \sigma_z} \leq \varepsilon_F < \epsilon_{N+1 \sigma_z}$.
With $r \sim 1$~nm we have roughly $N \sim 10 \div 100$. Hence, $N$ is large  but still much smaller than $m c r /\hbar \sim 10^{3}$, and (\ref{eq:validity-spect}) holds. In addition, the spacing between levels  $\sim 0.1$~eV i.e. $\sim 1000$~Kelvin is such that  the zero temperature approximation  is still valid at  room temperature.

To find the operator of the spin current we use again the standard procedure based on the time derivative of   the spin density observable $\vec{s}$, see e.g. \cite{Zulicke, Splettstoesser, Sun:2005} (other approaches are also possible, see e.g. \cite{Johal})
\begin{equation}
\frac{d}{d t} \vec{s} \equiv \frac{d}{d t} \Psi^\dag \widehat{\vec{s}} \Psi = \frac{1}{i \hbar}\Big[ \Psi^{\dag} \widehat{\vec{s}} \widehat{H}\Psi - (\widehat{H}\Psi)^\dag \widehat{\vec{s}} \Psi \Big],
\end{equation}
where $\Psi$ is the wave function. Using (\ref{eq:hamiltonian-def})--(\ref{eq:H-partial}), we obtain in the coordinate representation (recall that in our case the coordinate is the winding angle $\theta$) the continuity equation
\begin{equation}
\frac{d}{d t} s_i = -\frac{1}{r}\frac{\partial}{\partial \theta} \Psi^\dag \hat{J}_{\theta, i} \Psi
\label{eq:continuity}
\end{equation}
where $\hat{J}_{\theta, i}$ is the $i$th component of the spin 1d current operator circulating in the ring
\begin{equation}
\hat{J}_{\theta, i} = \frac{1}{2}\Big\{\frac{\hat{p}_\theta}{m r} - \frac{\hat{s}_z \hat{p}_\theta^2}{m^3 r^3 c^2} , \hat{s}_i\Big\} = \frac{\{\hat{v}_\theta, \hat{s}_i\}}{2} \ .
\end{equation}
Here $\{,\}$ is the anticommutator and $\hat{v}_\theta$ can be viewed as the spin velocity.

Using  the spectrum (\ref{eq:spectrum}) and the eigenstates (\ref{eq:H-states}), we find that in the leading $N^{-1}$ approximation the only non-vanishing component of the spin  current density is that in the  $z$ direction
\begin{multline}
\bar{j} = \frac{1}{r}\sum_{n=-N}^{N} \sum_{\sigma_z=\pm 1} \Psi^\dag_{n, \sigma_z}\hat{J}_{\theta, z} \Psi_{n, \sigma_z} \approx \\ \sum_{n=-N}^{N} \sum_{\sigma_z=\pm 1} \Psi^\dag_{n, \sigma_z}\frac{\hbar^2 \hat{\sigma}^2_z \hat{p}_\theta^2}{4 m^3 r^4 c^2} \Psi_{n, \sigma_z} \approx \\ \frac{1}{3} \frac{\hbar^4}{m^3 r^4 c^2} \left( \frac{r \sqrt{2 m \varepsilon_F} }{\hbar} \right)^3
\label{eq:persistent-current}
\end{multline}
The current is protected by the $
T$-invariance. Indeed, the current dissipation  requires the backscatterings without spin-flip, which
are impossible since the higher winding numbers are too distant (remember that the energy level spacings are $\sim 100 \div 1000$ K). Besides, all the states below the Fermi energy are  occupied. As a result, the backscatterings with spin-flip contribute
only to the same spin current.

An important fact is that the whole spectrum (\ref{eq:spectrum}) -- (\ref{eq:fermilevel})  contributes to (\ref{eq:persistent-current}) (except the levels with zero winding number). Note that one can rewrite (\ref{eq:persistent-current}) as
\begin{equation}\label{jbar}
\bar{j} \sim  \frac{8\pi}{3} \frac{\hbar}{m r} N \frac{\varepsilon_F}{m c^2}\frac{\hbar}{2}\frac{1}{2 \pi r}  \sim \hbar  \Big(\frac{v_F}{c}\Big)^2  \frac{v_F}{r} \
\end{equation}
One obtains an estimate  for  $\bar{j}$  as a constant times the velocity quantum uncertainty $\delta v \sim \hbar/m r$, the Fermi level number $N$, the relativistic factor ${\varepsilon_F}/({m c^2})$, the spin momentum ${\hbar}/{2}$, divided by the length of the ring $2\pi r$. Thus, the spin current  results from the combination of geometric, quantum and relativistic effects.

It follows also from (\ref{jbar}) that the current  density   is decays in $r$,  so that it could become negligible at a macroscopic scale. If, however, we consider a
a "metamaterial", i.e., a macroscopic sample paved by a big number of nanorings, then the resulting  spin current  can be quite large (a sample area of the order of $1$~mm$^2$ can contain up to $10^{12}$ of such nanorings).

It is known that the current of magnetic moment can produce an electric field decaying as $R^{-3}$ in the distance $R$ from the current  or  as $R^{-2}$ in the case of spin rotation \cite{Sun:2004,Sun:2005}).
This field $\vec{\mathcal{E}}$ can be estimated
by using  Lorentz force in the rest frame of the spin, i.e. via the transformation rules for a magnetic field in the rotating reference frame
\begin{equation}\label{eq:EBB}
\vec{\mathcal{E}} \approx - (\dot{\vec{\theta}} \times \vec{R})\times \vec{B} \ .
\end{equation}
Here $\vec{B}$ is the magnetic field induced by the electron magnetic moment in the rest frame and it is assumed that $R \gg r$  and $\dot{\theta} \ll c / R$, i.e., that the distance from the ring is much larger than its size and the rotation is slow enough.  In fact the validity of the formula is a rather delicate issue (see e.g. the review \cite{McDonald} and references therein), but we believe that the formula provides a correct order of magnitude in our estimates.  Since the magnetic moment of an electron with spin $s_z$ is  $g \mu_B s_z/ \hbar$, where $g\sim 2$ is the gyromagnetic factor and $\mu_B=e\hbar/(2m)$ is the Bohr magneton, the magnetic field induced by $s_z$ in the rest frame is (cf. \cite{Sun:2004})
\begin{equation}\label{eq:BBB}
\vec{B} \approx  -\frac{g \mu_0 \mu_B s_z}{2 \pi \hbar} \frac{\partial}{\partial \vec{R}} \frac{Z}{R^3}, \ Z=\vec{e}_z \cdot\vec{R}
\end{equation}
Replacing here $\dot{\theta} s_z$ by $2\pi \bar{j}$ and using (\ref{eq:persistent-current})-(\ref{eq:BBB}), we obtain the electric field in the laboratory reference frame
\begin{equation}
\vec{\mathcal{E}} \sim  \frac{\mu_0 g \mu_B }{3 r c^2} \left(\frac{ 2 \varepsilon_F}{m} \right)^{\frac{3}{2}} \bigg(\frac{\vec{R} + 2 Z \vec{e}_z}{R^3} - 3 \frac{Z^2\vec{R}}{R^5} \bigg)
\label{eq:vec-E}
\end{equation}
This leads to
\begin{equation}\label{eq:ET}
\mathcal{E} \sim \frac{4 \mu_0 \mu_B}{3 r R^2} \frac{\varepsilon_F}{m c^2} v_F 
\end{equation}
so that $\mathcal{E} \sim 10^{-2} \div 1$~V~m$^{-1}$ at the nanoscale ($R \sim 1 \div 10$~nm),  i.e. it is comparable or bigger than the electric field $\sim 10^{-2}$~V~m$^{-1}$ obtained  \cite{Sun} at a distance $5$ nm from a Rashba ring. It is also worth noting that
the electric field  due to Thomas precession may exist in any (not necessarily metallic) nanoparticle which confines  magnetic moments in motion.
Note also that (\ref{eq:continuity})-(\ref{eq:ET}) have been derived for a constant curvature: we believe, however, that the same conclusions can essentially  be reached in the case of a non constant  curvature.

Let us finally compare the electric force $F_T=e\mathcal{E}$ acting on the charge carrier due to Thomas spin precession   and the electric force $F_C$  due to the van der Waals-Casimir effect, the  only known so far electric force in metallic devices under the conditions of equilibrium and neutrality. According to \cite{Emig:2007}
\begin{equation}\label{eq:EC}
F_C \sim \frac{7 c_0}{\pi} \frac{\hbar c r^6}{R^8}, \\ c_0 = \frac{143}{16}\ \ R \gg r
\end{equation}
It follows then from (\ref{eq:ET}) and (\ref{eq:EC})
\begin{equation}
\frac{F_T}{F_C} \sim \frac{4 \pi}{ 21 c_0}\frac{e \mu_0 \mu_B v_F R^6}{\hbar c r^7} \frac{\varepsilon_F}{m c^2}
\label{eq:force-ratio}
\end{equation}
Let $R_0$ be the distance where  the Thomas force and the van der Waals-Casimir force  are equal:  from (\ref{eq:force-ratio}) one has
\begin{equation}
R_0 \sim \Big(\frac{m c^2}{\varepsilon_F}\frac{\hbar c r}{e \mu_0 \mu_B v_F}\Big)^\frac{1}{6} r,
\end{equation}
so that taking $r \sim 1$~nm one gets $R_0 \sim 10^2$~nm. In other words,  Thomas precession forces   may dominate  van der Waals-Casimir forces for distances from $10^2$nm and larger. It has already been mentioned above that using "metamaterials" (pavements of macroscopic devices by nanoparticles) one can obtain  forces proportional to the  area of the macroscopic sample.

\section{On a possible experiment on the anomalous magnetic moment of the photon}
We have so far considered  electrons because they possess a non-zero magnetic moment. On the other hand, photons have zero  magnetic moment despite their spin being 1. Nevertheless, there are theoretical arguments (see e.g. \cite{Anomal1:2010,
Anomal2:2010} and references therein) according to which the photon has an  anomalous magnetic moment $\mu_\gamma$. It is clear, however, that any experimental proof of this fact has to be fairly sophisticated. For instance, the use of  inhomogeneous magnetic fields for direct measurements would require extremely strong fields in view of a very large $c$ and a very small $\mu_\gamma$.

We would like  to point out that the appearance of an electric field from the photon magnetic moment due to  Thomas precession could provide a way to check whether $\mu_\gamma \neq 0$.

\begin{figure} [!htbp]
\center
{\subfigure[$\mathcal{E} \neq 0$]{\includegraphics[width=40mm]{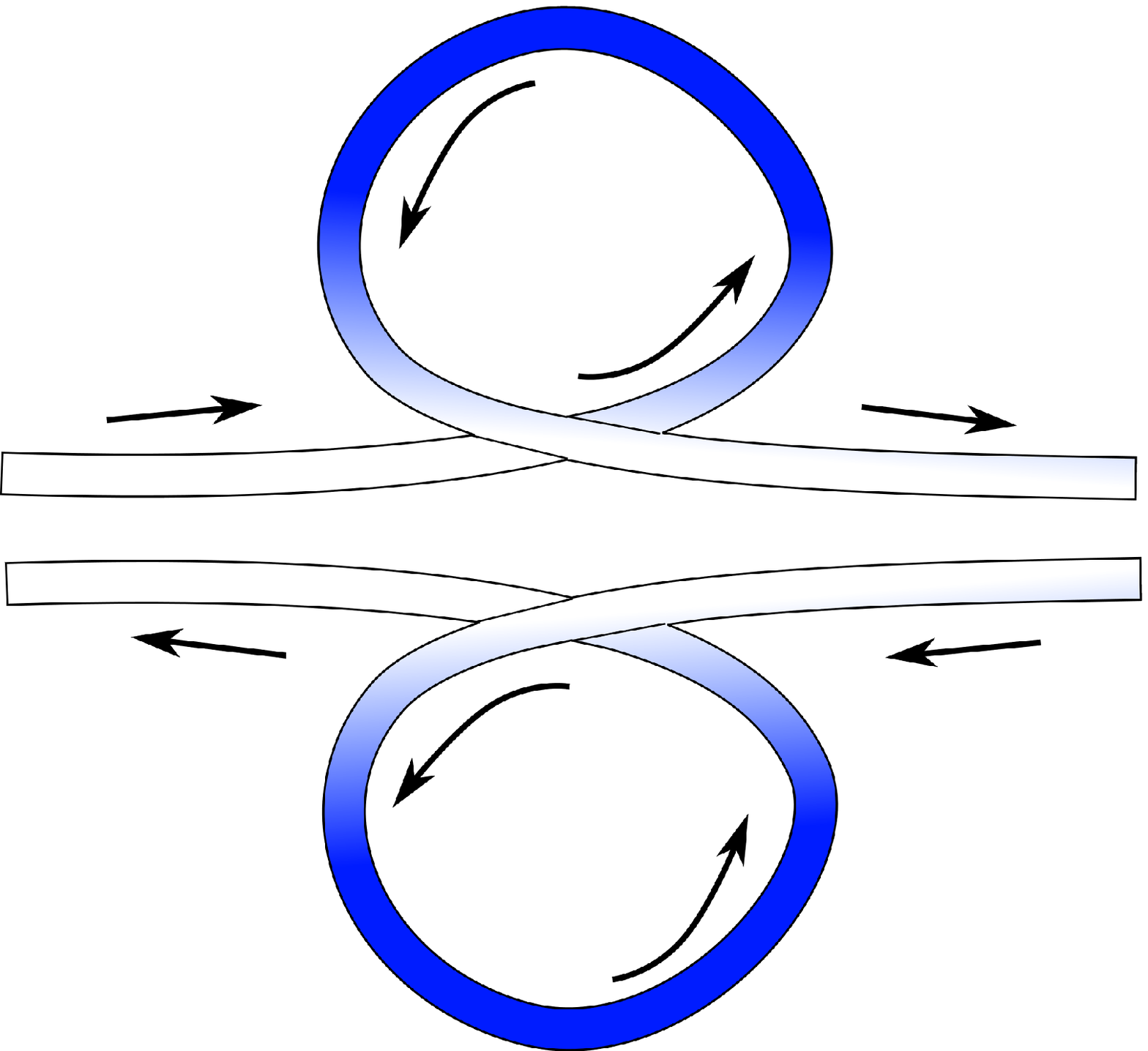}
\label{figure:lightguide1}}}
{\subfigure[$\mathcal{E} = 0$]{
\includegraphics[width=40mm]{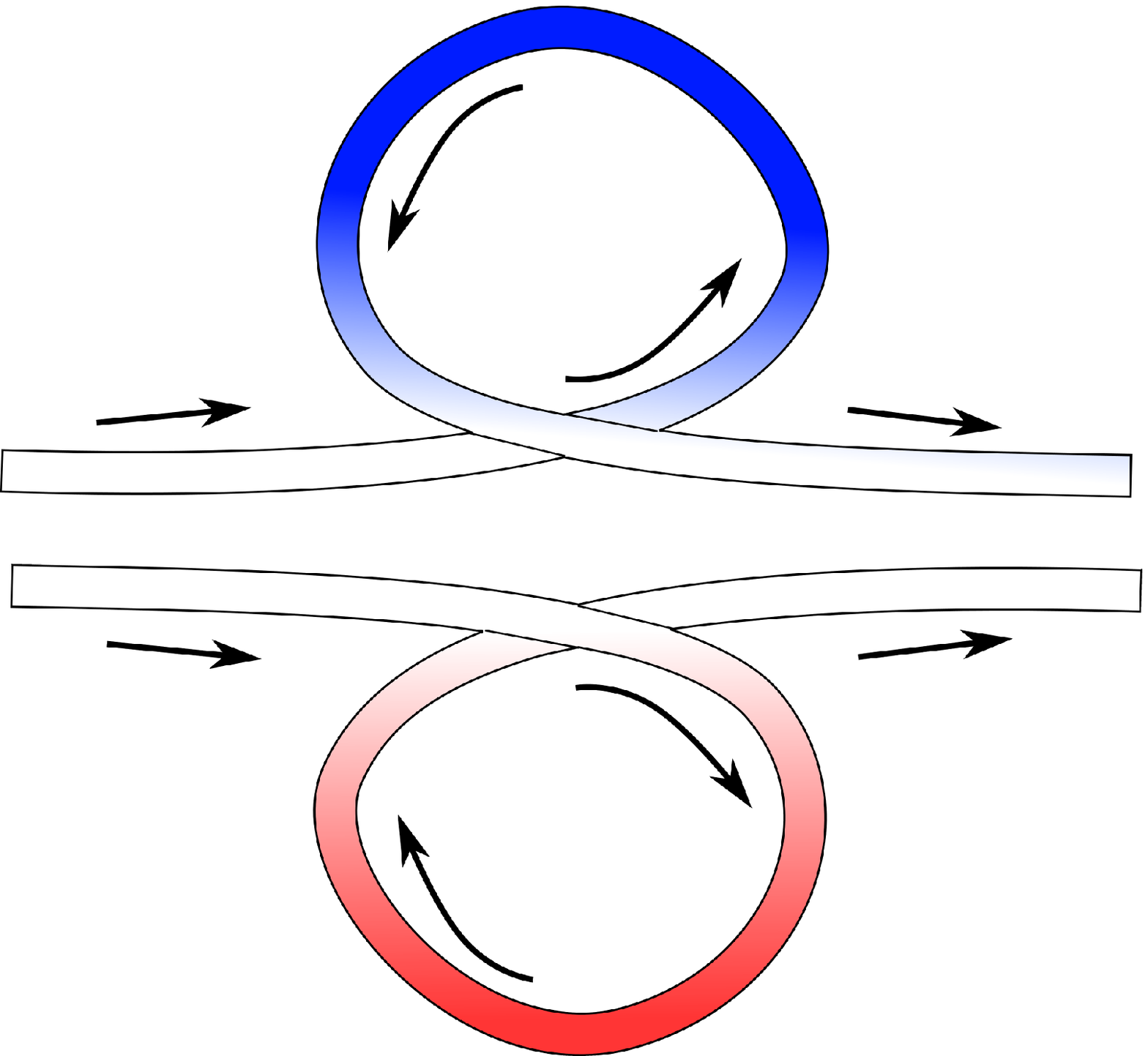}
\label{figure:lightguide2}}}
\caption{\subref{figure:lightguide1} -- light guide with nonzero spin current due to "additive" winding.
\subref{figure:lightguide2} -- light guide with "destructive" winding. The blue and red colors denote opposite spin precessions.}
\label{figure:lihgtguides}
\end{figure}

One has to consider normal (non-persistent) photon currents. To this end one
could use light guides with polarized light or recently found  100\% reflective materials \cite{TrappedLight} and measure the electric field from the confinement of
light in small volumes. In this case, however, it would be quite difficult to detect the contribution of the $\mu_\gamma$~current from other possible sources of the electric field. The idea is to use an initially unpolarized source of light and two light guides: one with a "constructive" winding of light  and the other with  a "destructive" (compensating)  winding, as shown in  Figure \ref{figure:lihgtguides}.

In this case one does not need to separate the contribution of the $\mu_\gamma$-current to the electric field from those stemming from  other sources, since the unique difference in the two light guides  is their winding directions, hence their Thomas currents, and so one has simply to  compare their electric fields. To estimate the effect we assume that the refraction coefficient $\mathfrak{n} = c/v$ of the light guide medium is $1.5 \div 2$ (optical glasses, crystals, etc), where $v$ is the speed of light  in the medium. We consider UV light with a wavelength  $\lambda \sim 100 nm$ and micro-meter windings with $r \sim 1 \mu m$. In this case the spin-orbit energy of the  Thomas precession is still less than the kinetic energy and we can use the formula
$$E_{s-o} \sim \frac{v^2}{2 c^2} s \dot{\theta} \sim \frac{\hbar c}{2 \mathfrak{n}^3 r} \ll \frac{\hbar c}{\lambda} $$
which doesn't contain $\mu_\gamma$. Hence, the effective magnetic field
$$
B \sim \frac{E_{s-o}}{\mu_{\gamma}} \sim \frac{\hbar c}{2 \mathfrak{n}^3 r \mu_B} \frac{\mu_B}{\mu_\gamma} \sim {10^5\over 2} \frac{\mu_B}{\mu_\gamma} \ \text{Tesla} \ .
$$
acting on $\mu_\gamma$ is quite strong.

It follows that Thomas precession could be used in order to detect $\mu_\gamma$ even if it is very small. Indeed, the resulting  photonic spin current $j_\gamma$  can be estimated as
$$
{ j_\gamma} \sim \rho_\gamma \hbar \frac{E_{so}}{p} \sim \rho_\gamma \hbar \frac{\lambda c}{\mathfrak{n}^3 r}
$$
where $p=h/\lambda$ is the photon momentum and $\rho_\gamma$ is the 1d  photon density.
The electric field due to Thomas precession follows by repeating the same steps as in (\ref{eq:vec-E}) with $g \mu_B$   replaced by $\mu_\gamma$ and  $\bar j$ by $j_\gamma$
$$
\mathcal{E}_\gamma \sim \frac{\mu_0 \mu_\gamma j_\gamma}{\hbar R^2} \sim \mu_0 \rho_\gamma \mu_\gamma \frac{\lambda c}{\mathfrak{n}^3 r R^2}  \ .
$$
Therefore, $\mu_\gamma$ can be detected from the measurements of $\mathcal{E}_\gamma$ for a sufficiently dense array of curved light circuits.

We can compare the electric field $\mathcal{E}_\gamma$ obtained here and the electric field $\mathcal{E}$ derived above for electrons in a metal nanoring of the same radius
$$
\frac{\mathcal{E}_\gamma}{\mathcal{E}} \sim \rho_\gamma \lambda \frac{\mu_\gamma}{\mu_B} \Big(\frac{c}{v_F}\Big)^3\frac{1}{\mathfrak{n}^3}
$$
We see that a very small value of $\mu_\gamma/\mu_B$ can be in part compensated by the cube of the inverse relativistic factor $v_F/c$.

We have  presented above estimates for winding effects in  light guide experiments. Similar effects could as well show up in the winding of  diffuse light and the associated magnetic moment edge currents for the strip geometry considered in \cite{Ouvry1}.

\section{Conclusion}
We have shown that the Thomas precession resulting from the confined motion of  spins carriers can generate some specific persistent spin currents. For the sake of concreteness, we have considered conducting electrons in metallic nanorings. It was pointed out, however, that similar effects can be expected for other spin carriers like magnons, spinons, ions in gases, etc. The Thomas precession of magnetic moments can generate electric forces that are stronger than the usual van der Waals-Casimir forces in metallic samples. Hence, it seems certainly appropriate to take Thomas precession effects into account when dealing with metallic nanoparticles. It has also been suggested that the electric forces due to  Thomas precession can be a possible tool for  experimentally testing the existence (see \cite{Anomal1:2010, Anomal2:2010} and references therein) of a non-zero anomalous magnetic moment for the photon.

\section{Acknowledgment}
This work was partly supported by the French-Russian program "Dnipro 2013-2014" (CNRS -- NASU).
A.~Yanovsky would like to thank the Laboratory of Theoretical Physics and Statistical Models (LPTMS) of CNRS, University Paris Sud for hospitality during the initial stage of the work.

\end{document}